\documentclass[a4paper,11pt]{article}
\usepackage[T1]{fontenc}
\usepackage[utf8]{inputenc}
\usepackage{lmodern}
\usepackage{amssymb,amsmath,amsthm,amsfonts}
\usepackage{mathtools,mathrsfs}
\usepackage[english]{babel}
\usepackage{authblk}
\usepackage{indentfirst}
\usepackage{multirow,multicol}
\usepackage{booktabs,array,tabularx}
\usepackage{caption}
\usepackage{float}
\usepackage{fancyhdr}
\usepackage[numbers]{natbib}
\usepackage{xcolor,colortbl}
\usepackage{url}
\usepackage{microtype}
\usepackage{graphicx}
\usepackage{pgfplots}
\pgfplotsset{compat=1.18}
\usepackage[colorlinks=true,allcolors=blue]{hyperref}
\usepackage{bookmark}
\hypersetup{
 pdftitle={Local magnitude weighting for protein flexibility},
 pdfauthor={Wanying Bi, Jian Liu, Hongsong Feng},
 pdfsubject={Mathematical structure, B-factor modeling, and computational scaling},
 pdfkeywords={local magnitude, protein flexibility, B-factor, kernel weighting}
}
\usepackage{placeins}
\newtheorem{theorem}{Theorem}[section]

\newtheorem{proposition}[theorem]{Proposition}

\theoremstyle{definition}
\newtheorem{definition}[theorem]{Definition}

\newcommand{\R}{\mathbb{R}}
\newcommand{\one}{\mathbf{1}}

\newcommand{\Ang}{\text{\AA}}

\newcommand{\Mag}{\mathcal{M}}
\newcommand{\norm}[1]{\left\lVert #1\right\rVert}

\title{Protein B-factor modeling with local magnitude}
\author[1]{Wanying Bi}
\author[1]{Jian Liu$^\dag$}
\author[2]{Hongsong Feng\thanks{Corresponding authors: liujian@cqut.edu.cn, hfeng2@charlotte.edu}}
\affil[1]{School of Mathematical Sciences, Chongqing University of Technology, Chongqing 400054, China}
\affil[2]{Department of Mathematics and Statistics, University of North Carolina at Charlotte, Charlotte, NC 28223, USA}

\makeatletter
\renewcommand*{\@fnsymbol}[1]{\ensuremath{\ifcase#1\or\dagger\or *\or\ddagger\or\mathsection\or\mathparagraph\or\|\or **\or\dagger\dagger\or\ddagger\ddagger\else\@ctrerr\fi}}
\makeatother
\date{}

\begin{document}
\maketitle
\paragraph{Abstract}
Topological magnitude characterizes the effective size and spatial organization of metric spaces, offering a geometric approach with potential applications in molecular biology and materials science. Previous work has demonstrated its utility in modeling protein flexibility from structure, but global magnitude weighting requires dense linear systems whose computational cost increases with protein size. We introduce a local magnitude representation that retains residue-level structural information through nearest-neighbor environments. The construction is well posed and geometrically invariant, with central weighting distinguished from scalar neighborhood magnitude. Across 364 proteins, the mean within-protein fitting Pearson correlation coefficient (PCC) increases from 0.725 for global magnitude to 0.728 for local magnitude. Blind-prediction experiments on 345 proteins combine each representation with identical auxiliary descriptors and evaluate random forests and gradient boosting under three validation protocols. Local features improve the primary PCC in all six model--protocol comparisons, with a maximum absolute increase of 0.008 in protein-level ten-fold gradient boosting. For a fixed feature bank, local factorization costs $O(Nk^3)$ compared with $O(N^3)$ for global factorization, where $N$ is the number of C$_\alpha$ atoms and $k$ the neighborhood size. Timing measurements show a computational advantage of local magnitude on large structures. Within the tested single-seed configuration and distinct global and local scale banks, these results support local magnitude as a compact representation that preserves useful B-factor information while reducing feature-extraction costs for large proteins.


\clearpage
\tableofcontents
\clearpage

\section{Introduction}
Protein flexibility is central to molecular recognition, catalysis, and conformational adaptation \cite{frauenfelder1991,teilum2009}. A protein must retain a stable overall structure while permitting motions associated with function \cite{teilum2011}. Atomic displacement parameters obtained from crystallographic structures provide an accessible, although incomplete, experimental description of structural disorder. Predicting their B-factors from a known structure is therefore useful both as a modeling problem and as a test of structural representations.

Several complementary approaches describe protein motions and flexibility. Molecular dynamics follows trajectories generated by an interaction model \cite{mccammon1977}, whereas normal mode analysis characterizes near-equilibrium motions through collective vibrational modes \cite{go1983,levitt1985,ma2005}. Gaussian network models (GNM) represent residue contacts using an isotropic harmonic network \cite{bahar1997}, and anisotropic network models retain directional information \cite{atilgan2001}; coarse-grained and approximate formulations extend these approaches to collective conformational changes and residue-level fluctuations \cite{tama2001,li2002,yang2008,park2013}. Graph-theoretic rigidity analysis and stochastic models provide additional structure-based descriptions of protein flexibility \cite{jacobs2001,xia2013stochastic}. Flexibility--rigidity indices (FRI), including fast, anisotropic, and multiscale formulations, describe atomic environments through distance-dependent correlation kernels and combine length scales to model protein thermal fluctuations \cite{xia2013fri,opron2014,opron2015}.

Geometric and topological methods characterize protein flexibility through atom-specific persistent homology \cite{bramer2020} and evolutionary homology on coupled dynamical systems \cite{cang2020}. Multiscale Gauss linking integrals (mGLI) \cite{shen2024mgli} and Jones-polynomial descriptors \cite{song2025jones} encode local curve entanglement for B-factor prediction, while multiscale differential geometry learning uses mean and Gaussian curvatures for fitting and blind prediction \cite{feng2025mdg}. Weighted persistent homology has also been applied to RNA flexibility \cite{pun2020}. Together, these methods connect local molecular structure to flexibility across multiple scales.

Magnitude measures the size of finite metric spaces through their distance structure \cite{leinster2013}, with related graph and homological constructions \cite{leinster2019,hepworth2017,bi2024hypergraph}. Persistent magnitude homology captures multiscale geometric and topological information using weighted barcodes \cite{bi2026persistent}, a construction distinct from numerical magnitude weighting. Applications include image edge detection using magnitude vectors \cite{adamer2024} and carborane property analysis using persistent magnitude \cite{bi2024carboranes}. The image and protein formulations share the underlying weighting construction but use different metric representations. For protein B-factor modeling, inverse row sums of a similarity matrix define C$_\alpha$-level weights across kernel parameters \cite{bi2025}. Whereas scalar magnitude sums these weights, the vector retains residue-level information, motivating the center-specific local descriptor studied here.

A local construction offers a different balance between descriptive detail and computational cost. Each atom is represented through a bounded neighborhood, so dense linear algebra can be limited to small matrices. This also separates local packing from changes elsewhere in a large structure. However, localization changes the descriptor: a local weighting is generally not equal to the corresponding component of the global weighting. Moreover, solving many small systems incurs overhead, and cannot be assumed to be faster for every protein size.

This study makes three contributions. First, it defines local magnitude and central weighting and establishes the well-posedness and geometric invariance of the resulting descriptor. Second, it gives an explicit C$_\alpha$-only feature construction and a worked protein example, while clarifying the interpretation of flexibility and rigidity. Third, it evaluates the resulting fixed feature bank using the archived data and evaluation protocols of the global magnitude, and compares computational costs on five named proteins. The emphasis is on a transparent comparison against a baseline matched for data, model, and evaluation protocol, including cases where localization does not improve performance.

The study addresses three related questions: how to define a local weighting with a clear mathematical interpretation; whether a fixed, compact feature bank retains useful B-factor information; and at what structural sizes its computational cost becomes favorable. Section~\ref{sec:methods} develops the theory and modeling procedure. Section~\ref{sec:results} compares fitting, prediction, and timing results, and examines individual protein examples. Section~\ref{sec:discussion} discusses their implications and limitations.

\section{Theory and methods}
\label{sec:methods}

\begin{figure}[htbp]
\centering
\includegraphics[width=\linewidth]{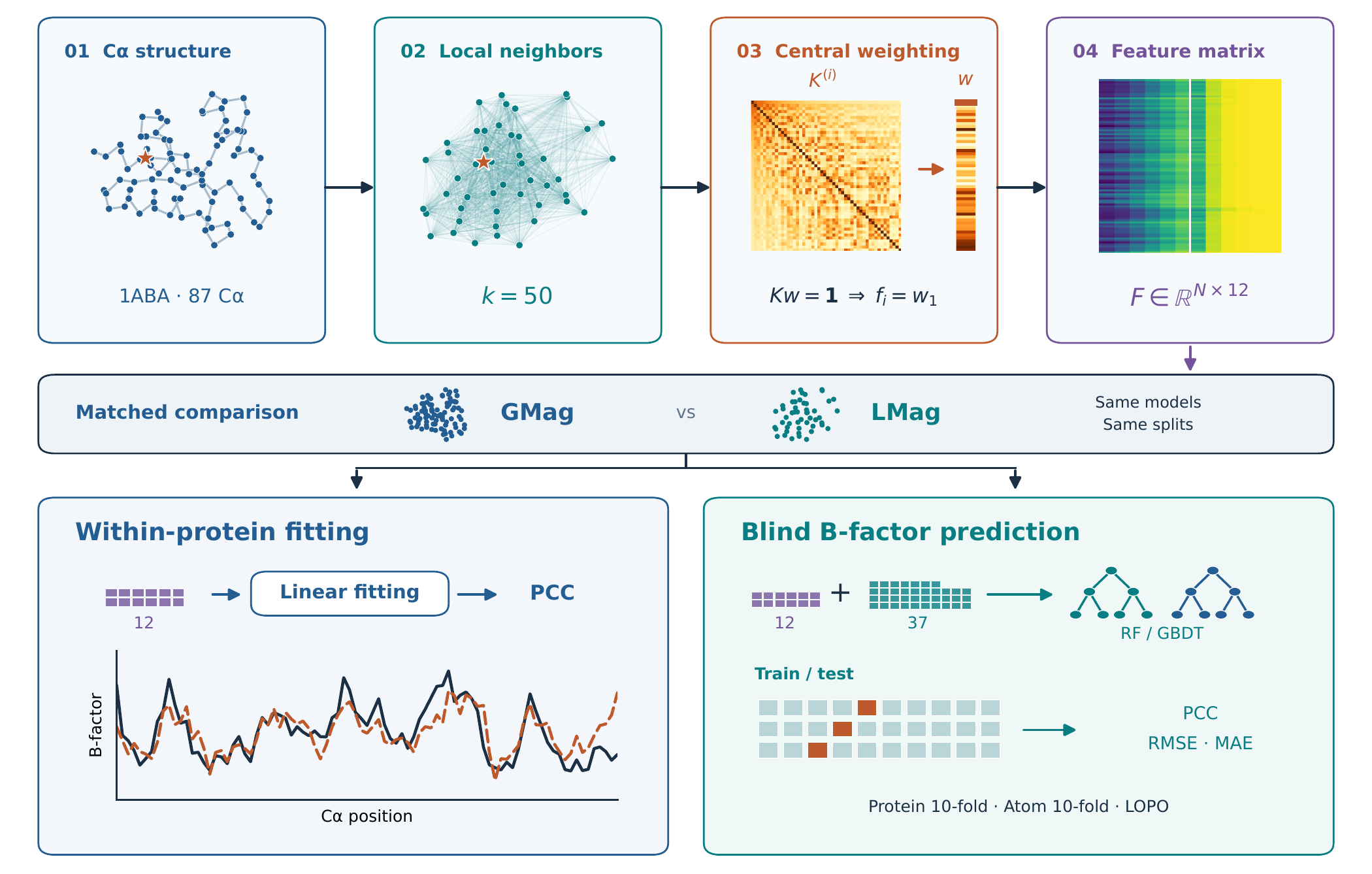}
\caption{Local magnitude workflow: from C$_\alpha$ geometry to B-factor fitting and blind prediction. Stars mark neighborhood centers. The 1ABA profile shows experimental values (solid) and the local fit (dashed); orange blocks indicate held-out targets in the schematic validation diagram.}
\label{fig:method-overview}
\end{figure}

Figure~\ref{fig:method-overview} summarizes the proposed workflow, from C$_\alpha$ coordinates to local central-weighting features and B-factor modeling. The same geometric representation supports within-protein fitting and blind prediction, with auxiliary descriptors added only in the latter setting.

\subsection{Local magnitude}
\label{sec:theory}
Let $X=\{x_1,\ldots,x_N\}\subset\R^3$ be distinct points and let $d_{ij}=\norm{x_i-x_j}_2$. For a center $i$, define $S_i$ to contain the center and its $m_i-1$ nearest other points, where $m_i=\min(k,N)$. An alternative definition uses a radius $R$ and includes all points with $d_{ij}\leq R$. The experiments below fix $k=50$. Ties are resolved by the archived row index; the center is always placed first.

For $\alpha>0$ and $p\in\{1,2\}$, define
\begin{equation}
 K^{(i)}_{ab}(\alpha,p)
 =\exp\!\left(-\alpha\norm{x_{s_a}-x_{s_b}}_2^p\right),
 \qquad s_1=i,\quad s_a\in S_i.
 \label{eq:kernel}
\end{equation}
Coordinates are expressed in angstroms, so $\alpha$ has units $\Ang^{-p}$. An associated length scale is $\ell=\alpha^{-1/p}$, for which $K(d=\ell)=e^{-1}$. Equal numerical $\alpha$ values for $p=1$ and $p=2$ therefore do not specify equal length scales.

\begin{definition}
For an invertible neighborhood kernel, let $w^{(i)}$ satisfy $K^{(i)}w^{(i)}=\one_{m_i}$. The \textit{neighborhood magnitude} $\Mag_i$ and its \textit{central weighting component} $f_i$ are defined as
\begin{equation}
 \Mag_i=\one_{m_i}^{T}w^{(i)},\qquad
 f_i=w^{(i)}_1.
 \label{eq:definitions}
\end{equation}
In the context of protein flexibility analysis, we refer to $f_i$ the \textit{local magnitude flexibility index} (LMFI).
\end{definition}

For $p=1$, $\Mag_i$ is the usual magnitude of the finite metric space with distance $\alpha d$. For $p=2$, squared Euclidean distance need not satisfy the triangle inequality. The same expression is then a positive-definite-kernel magnitude, not automatically the magnitude of an ordinary metric space. This distinction does not prevent solving \eqref{eq:definitions}, but matters when importing metric-space theorems.

\begin{proposition}
For distinct Euclidean points, $\alpha>0$, and $p=1$ or $p=2$, $K^{(i)}$ is strictly positive definite. Thus the local weighting is unique and $\Mag_i>0$.
\end{proposition}
\begin{proof}
The Gaussian kernel ($p=2$) is strictly positive definite on distinct Euclidean points. One way to see this is through its strictly positive Fourier density: its quadratic form is an integral of $|\sum_a c_a e^{\mathrm{i}\omega\cdot x_{s_a}}|^2$ against a positive Gaussian measure, which is nonzero for any nonzero coefficient vector. For $p=1$, the identity
\begin{equation}
 e^{-\alpha r}=
 \frac{\alpha}{2\sqrt{\pi}}
 \int_0^\infty t^{-3/2}
 e^{-\alpha^2/(4t)}e^{-t r^2}\,\mathrm{d}t
 \label{eq:mixture}
\end{equation}
expresses the kernel as a positive mixture of strictly positive-definite Gaussian kernels. This is consistent with the classical kernel results of Schoenberg \cite{schoenberg1938}. Positive definiteness implies invertibility, and $\Mag_i=\one^T(K^{(i)})^{-1}\one>0$.
\end{proof}

Positive definiteness does not imply componentwise-positive weightings. For example, consider three collinear points at $-1,0,1$, with the middle point as center, $p=2$, and $q=e^{-\alpha}$. Symmetry gives
\begin{equation}
 f_{\mathrm{center}}=
 \frac{1+q^4-2q}{1+q^4-2q^2}.
 \label{eq:negative}
\end{equation}
The denominator is $(1-q^2)^2>0$, but the numerator is negative for, for example, $\alpha=0.1$. 

The construction depends on pairwise distances and is invariant under translations, rotations, and reflections. It is equivariant under relabeling when neighbor selection is unchanged, in particular when there are no ties at the neighborhood boundary. At a tied cutoff, the row-index tie rule is a reproducibility convention rather than an intrinsic geometric choice. Uniform coordinate scaling by $c>0$ is compensated by replacing $\alpha$ with $\alpha/c^p$. As $\alpha\to\infty$, every fixed neighborhood kernel approaches the identity, so $f_i\to1$ and $\Mag_i\to m_i$. Very short kernel length scales consequently reduce feature contrast.

The central weighting gives a field of center-specific descriptors rather than a single scalar for the protein, as does the global weighting vector. Localization additionally bounds the size of each factorization independently of $N$ when $k$ is fixed. It does not recover magnitude homology, nor is a local weight guaranteed to approximate a global one uniformly. To make this last point explicit, partition a global kernel as
\begin{equation}
 K_X=\begin{pmatrix}A&B\\B^T&C\end{pmatrix},\qquad
 K_X\binom{g}{h}=\binom{\one}{\one},
 \qquad u=A^{-1}\one,
\end{equation}
where $A$ corresponds to $S_i$. Then
\begin{equation}
 g-u=-A^{-1}Bh,\qquad
 |g_1-u_1|\leq\norm{e_1^TA^{-1}}_2\norm{B}_2\norm{h}_2.
 \label{eq:localerror}
\end{equation}
Weak distant similarity alone is insufficient: conditioning and the outside weighting also affect the localization error. Appendix~\ref{app:stability} gives a corresponding perturbation bound for fixed neighborhoods.

\subsection{Local weighting as a flexibility descriptor}
\label{sec:rigidity}
We represent each retained residue by its C$_\alpha$ coordinate, without using B-factors to construct neighborhoods or kernels. The term flexibility index in $f_i$ designates a geometric predictor, not a derived mechanical response law. In particular, there is no theorem that $f_i$ must increase with the experimentally measured B-factor for every residue.

A related rigidity convention uses the reciprocal of the flexibility index. The convention introduced in \cite{bi2025} can be localized as
\begin{equation}
 r_i^{\mathrm{rec}}=1/f_i\quad\text{when }f_i\ne0.
 \label{eq:reciprocal}
\end{equation}
It is a positive rigidity-like score only on the domain $f_i>0$; near zero it is unstable, and for negative $f_i$ it is not a positive stiffness. The conventional kernel-density rigidity score of FRI is instead $r_i^{\mathrm{dens}}=\sum_{j\in S_i}K_{1j}^{(i)}>0$, with flexibility $1/r_i^{\mathrm{dens}}$ \cite{opron2014}. The inverse row sum in \eqref{eq:definitions} generally differs from this reciprocal row sum.

The reported feature bank uses the central weights directly; reciprocal scores are not included. For a singleton neighborhood, the central weight and neighborhood magnitude both equal one. In larger neighborhoods, the central weight depends on the full arrangement of neighboring atoms and is used as a geometric input to the B-factor models.

\subsection{Feature extraction and a worked protein example}
\label{sec:example}
The fixed representation uses the ordered parameter bank
\begin{equation}
 \mathcal{P}=\{(\alpha,p):p\in\{1,2\},\quad
 \alpha\in\{0.1,0.2,0.3,0.4,0.5,0.6\}\}.
 \label{eq:bank}
\end{equation}
The six $p=1$ components precede the six $p=2$ components, each in increasing $\alpha$ order. For every center we select the neighborhood, form all of its pairwise distances, construct one kernel per parameter pair, and compute
\begin{equation}
 K^{(i)}=LL^T,\qquad Lz=\one,\qquad L^Tw^{(i)}=z.
 \label{eq:cholesky}
\end{equation}
Only the central component is retained as a feature. The implementation uses double precision and checks $\norm{K^{(i)}w^{(i)}-\one}_\infty\leq10^{-10}$. Duplicate coordinates or a failed factorization cause an error rather than silent diagonal regularization. Residual checks test numerical consistency; without a condition-number bound they are not a prescribed relative-error guarantee. No explicit matrix inverse is needed.

Consider the coordinate set of protein 1ABA, containing 87 C$_\alpha$ atoms. The first retained atom has coordinates $(19.791,50.519,58.057)\,\Ang$. Its 50-point neighborhood includes itself and extends to a farthest-center distance of $18.060648\,\Ang$. The first few selected coordinate-row indices are $0,1,30,70,29,2$, with distances $0,3.792428,4.793719,4.794559,5.195897,6.279256\,\Ang$, respectively. Indices refer to archived rows, not necessarily consecutive PDB residue numbers. This illustrates that spatial neighbors need not be adjacent in sequence.

At $p=1,\alpha=0.1$, the resulting $50\times50$ system gives $f_i=0.166045$ and $\Mag_i=3.807301$. The other parameter pairs produce Table~\ref{tab:example}. Repeating the process for all 87 centers yields an $87\times12$ feature matrix. The total neighborhood magnitude in the table is explanatory and is not a column of that feature matrix.

\begin{table}[htbp]
\centering
\caption{Worked example for the first retained C$_\alpha$ of 1ABA.
Each row uses the same 50-point neighborhood. LMFI is the feature used by the
prediction models; $\Mag_i$ is the separate neighborhood total.}
\label{tab:example}
\begin{tabular}{rrrr}
\toprule
$p$ & $\alpha$ & LMFI $f_i$ & $\Mag_i$\\
\midrule
1&0.1&0.166045&3.807301\\
1&0.2&0.275381&8.673247\\
1&0.3&0.399210&15.379421\\
1&0.4&0.530026&22.930001\\
1&0.5&0.653274&30.086374\\
1&0.6&0.756426&36.018802\\
2&0.1&0.712982&31.106769\\
2&0.2&0.928865&45.184618\\
2&0.3&0.984742&48.892088\\
2&0.4&0.996624&49.746225\\
2&0.5&0.999226&49.941532\\
2&0.6&0.999819&49.986471\\
\bottomrule
\end{tabular}
\end{table}

The neighborhood count and kernel length scale play different roles. Here the 50-neighbor radius is about $18\,\Ang$, whereas the $p=1$ e-folding lengths range from $10$ to $1.67\,\Ang$. For $p=2$ they range from $3.16$ to $1.29\,\Ang$. Consequently, including 50 atoms does not assign all 50 equally strong interactions, and fixing $k$ does not impose a fixed physical radius.

Table~\ref{tab:lengths} makes the scale interpretation explicit. Increasing $\alpha$ decreases the interaction length, while increasing $k$ extends the set of points participating in the solve. These are separate controls: changing $k$ at fixed $\alpha$ modifies the neighborhood geometry, whereas changing $\alpha$ at fixed $k$ changes the similarity assigned to its distances. The regular grid makes the construction easy to reproduce; it is not claimed to be an optimal or nonredundant basis for all proteins.

\begin{table}[htbp]
\centering\small
\caption{E-folding interaction lengths $\ell=\alpha^{-1/p}$ in angstroms
for the local feature bank. The numerical alpha values have units
$\Ang^{-1}$ for $p=1$ and $\Ang^{-2}$ for $p=2$.}
\label{tab:lengths}
\begin{tabular}{lrrrrrr}
\toprule
Numerical $\alpha$&0.1&0.2&0.3&0.4&0.5&0.6\\
\midrule
$\ell$, $p=1$&10.00&5.00&3.33&2.50&2.00&1.67\\
$\ell$, $p=2$&3.16&2.24&1.83&1.58&1.41&1.29\\
\bottomrule
\end{tabular}
\end{table}

The complete calculation can be implemented as the following sequence. First, identify the retained C$_\alpha$ records and preserve their mapping to the target rows. Second, build the spatial neighbor index once per structure and select at most 50 points for each center. Third, form the pairwise distance matrix within each selected neighborhood. Fourth, solve the 12 kernel systems by \eqref{eq:cholesky} and store the center component in its corresponding feature column. Finally, pass the resulting $N\times12$ array to the selected regression protocol. Targets enter only this last stage. Caching this array allows the same geometric calculation to be reused for different folds and regression models. Independent LU solves verified selected local weights, including the 1ABA example, with a maximum discrepancy of approximately $9.160\times10^{-16}$.

\subsection{B-factor modeling, data alignment, and evaluation}
\label{sec:evaluation}
For an isotropic displacement parameter $U_{\mathrm{iso}}$, the crystallographic convention is $B=8\pi^2 U_{\mathrm{iso}}$. If displacement is isotropic and $\langle\norm{u}^2\rangle$ denotes total three-dimensional mean-square displacement, then
\begin{equation}
 B=\frac{8\pi^2}{3}\langle\norm{u}^2\rangle.
\end{equation}
The distinction avoids ambiguity between a one-axis variance and the total displacement. B-factors also reflect static disorder, crystal packing, occupancy, experimental conditions, and refinement. They are therefore a proxy target, not an exhaustive description of solution-state dynamics.

\paragraph{Data provenance.}
We distinguish two evaluation settings: within-protein B-factor fitting (Task 1) and blind B-factor prediction (Task 2). The former fits and evaluates each protein using its own observed labels, whereas the latter evaluates predictions on held-out labels. We use the archived public inputs accompanying Bi et al.\ \cite{bi2025}. The within-protein fitting cohort contains 364 proteins and 78,403 genuine C$_\alpha$ records. The blind-prediction cohort follows the supplied implementation's exclusion of 3P6J, leaving 345 proteins and 74,527 target rows. The published article describes 346 proteins for this task; the code/data-aligned cohort used here has 345. We retain task-specific coordinate filtering and archived row order. Some supplied blind-prediction labels differ from the B-factor fields in its PDB files; we use the supplied label tables and do not replace them with those fields.

The exclusion criteria described by Bi et al. include unavailable secondary-structure assignments, known data problems, zero-valued target records, and mismatches between structural and descriptor rows. We inherit the supplied cohorts rather than downloading replacement structures and constructing new exclusion lists. In particular, the geometric representation contains only the retained C$_\alpha$ coordinates: hydrogens, side-chain atoms, waters, and ligands are not added to the 50-point neighborhoods. Missing residues are not imputed. These choices keep the point sets aligned with the archived labels, while limiting what the geometry can describe.

Local features for blind prediction are recomputed from the corresponding coordinates, rather than reusing neighborhoods from the within-protein fitting cohort. The global baseline features for blind prediction and 37 auxiliary features are read from the author's provided tables. The global bank for within-protein fitting was recomputed on the C$_\alpha$ coordinates, using the original paper's kernel parameters. Table \ref{tab:cohorts} records the actual usable cohort sizes. Named Small, Medium, and Large subsets are inherited lists, not new bins defined by the present atom counts. They overlap the Superset and must not be summed with it. In addition, $k=50$ covers the full protein for 40 proteins in the fitting cohort and 35 in the blind-prediction cohort, so the construction is not strictly local in those cases.

\begin{table}[htbp]
\centering
\caption{Actual cohort sizes used in this study. C$_\alpha$ counts are retained
rows, not total heavy-atom counts. Subsets are author-supplied membership lists.}
\label{tab:cohorts}
\begin{tabular}{lrrrr}
\toprule
&\multicolumn{2}{c}{Within-protein fitting}&\multicolumn{2}{c}{Blind prediction}\\
\cmidrule(lr){2-3}\cmidrule(lr){4-5}
Subset&Proteins&C$_\alpha$&Proteins&C$_\alpha$\\
\midrule
Small&33&943&30&899\\
Medium&37&3342&37&3341\\
Large&35&5515&34&5390\\
Superset&364&78403&345&74527\\
\bottomrule
\end{tabular}
\end{table}

\paragraph{Within-protein B-factor fitting.}
For each protein separately, an ordinary least-squares model with an intercept fits all of that protein's target values:
\begin{equation}
 \widehat B_i=\beta_0+\sum_{q=1}^{12}\beta_q f_{iq}.
 \label{eq:ols}
\end{equation}
Feature columns are centered and divided by their Euclidean norms to improve numerical stability, with near-zero columns set to zero. The centered least-squares solve uses a relative rank cutoff of $10^{-10}$. The intercept restores the target mean. No auxiliary features are used. This is an in-sample representation test, not prediction for unseen labels. In particular, very short proteins can be nearly interpolated by a model with 12 columns and an intercept.

\paragraph{Blind B-factor prediction with auxiliary features.}
Let $a_i\in\R^{37}$ contain the supplied auxiliary descriptors. We compare
\begin{equation}
 z_i^{\mathrm{local}}=(f_i^{(1)},\ldots,f_i^{(12)},a_i),\qquad
 z_i^{\mathrm{global}}=(g_i^{(1)},\ldots,g_i^{(12)},a_i),
\end{equation}
both of dimension 49. The auxiliary vector contains 20 amino-acid indicators, seven secondary-structure indicators, and ten numeric descriptors: $\phi$, $\psi$, solvent-accessible area, resolution, R-value, heavy-atom count, occupancy, and three packing-density quantities. The target B-factor is excluded from the inputs. Thus the local magnitude block is geometry-only, but the complete blind-prediction model is not.

The original workflow obtains secondary structure, backbone angles, and accessible surface area through STRIDE \cite{heinig2004}. Its packing descriptors distinguish distances below $3\,\Ang$, between $3$ and $5\,\Ang$, and beyond $5\,\Ang$ \cite{bi2025}. We use the supplied encoded values without regenerating them. One-hot encoding expands the amino-acid and secondary-structure categories, so the 37 input columns should not be confused with a count of 37 distinct physical observables. The experimental resolution and R-value also carry information about the structure determination; the complete predictor therefore combines local geometry with residue annotations and experimental context.

Random forests (RF) \cite{breiman2001} and gradient-boosted decision trees (GBDT) \cite{friedman2001} use 1000 trees and seed zero. Even with the same random seed, the training results still vary across runs because the training set is sampled randomly. Both input versions use identical rows, folds, and hyperparameters, without feature or target standardization. Appendix~\ref{app:models} lists the settings. Three evaluation protocols are used:
\begin{enumerate}
\item \emph{Protein ten-fold:} shuffle the ordered protein list and hold out one of ten protein groups. No protein appears in both parts of a fold.
\item \emph{Atom ten-fold:} shuffle the target rows and hold out one of ten row groups. Other residues from a test protein can be in training.
\item \emph{Leave-one-protein-out (LOPO):} fit on all other proteins and predict every retained residue of the omitted protein.
\end{enumerate}
Protein-level separation is not sequence-family or homology-cluster separation. Atom ten-fold is particularly not an unseen-protein test. The known structures are used to form label-free geometric features in all protocols.

\paragraph{Scores and parameter selection.}
We compute Pearson correlation (PCC), root-mean-square error (RMSE), and mean absolute error (MAE):
\begin{align}
 \rho(y,\widehat y)&=
 \frac{\sum_j(y_j-\bar y)(\widehat y_j-\overline{\widehat y})}
 {\sqrt{\sum_j(y_j-\bar y)^2\sum_j(\widehat y_j-\overline{\widehat y})^2}},\\
 \operatorname{RMSE}&=\sqrt{n^{-1}\sum_j(y_j-\widehat y_j)^2},\qquad
 \operatorname{MAE}=n^{-1}\sum_j|y_j-\widehat y_j|.
\end{align}
Within-protein fitting PCC is averaged equally over proteins. For the two ten-fold protocols, the reported PCC averages the ten test-fold PCCs; LOPO averages the per-protein PCCs. Errors are computed over pooled target rows, in $\Ang^2$. Fold-mean, protein-mean, and pooled PCC are different statistics and are not interchanged. We also compare absolute error with a predictor that returns the training-fold mean B-factor for every test row.

The final bank is a uniform configuration, not a protein-specific choice. Nevertheless, its neighborhood and scale choices were informed by earlier within-protein fitting exploration on proteins overlapping the blind-prediction cohort. No further tuning on the blind-prediction evaluation is performed in the fixed comparison. The paper reports repeated random evaluations; our comparison uses one fixed model seed. Saved predictions were checked for label alignment and test-set coverage, and the reported scores were independently recomputed.

\subsection{Computational complexity and example timing protocol}
\label{sec:complexity}

Let $Q=12$ be the number of kernels. An efficiently implemented global method forms a distance matrix once and performs $Q$ dense Cholesky factorizations and solves. Its leading time is $O(QN^3)$, with $O(N^2+QN)$ working/output storage when kernels are processed sequentially. One global solve returns all $N$ weighting components. Explicit inversion, as in the original reference code, is unnecessary for this purpose.

The present local implementation performs $Q$ factorizations per center, giving solve cost
\begin{equation}
 O\!\left(Q\sum_{i=1}^{N}m_i^3\right)\leq O(QNk^3),
 \label{eq:complexity}
\end{equation}
plus neighbor searches and $O(\sum_i m_i^2)$ pairwise-distance work. Here, $k=50$ is the maximum number of neighboring atoms retained for each center. The current implementation processes one neighborhood at a time and stores the output, giving approximately $O(N+QN+k^2)$ memory apart from implementation constants. For fixed $k$, factorization cost is linear in $N$, but repeated Python calls and many small systems can dominate at small $N$. A radius-based neighborhood does not bound $m_i$ and therefore does not guarantee this scaling. If $k\geq N$, the unoptimized center-by-center code redundantly solves full systems.

We time five named proteins: 1ABA, 3G1S, 1KMM, 1H6V, and 1QKI, as shown in Table \ref{tab:timing}. Both methods use Cholesky and the same alpha bank \eqref{eq:bank}; local neighborhoods have $k=50$. Timed calls compute all 12 features for every C$_\alpha$ atom, rather than one atom or a single scalar magnitude. They include input checks, distances, kernels, factorization, solves, and residual checks; local calls also include neighbor selection. File loading, correctness comparisons, plots, regression, and model training are excluded. Methods run serially with one BLAS thread, after a small-system warm-up. Their order is reproducibly randomized within each protein/repeat. Three repeats are retained, including raw measurements. These are a few scaling examples, not a benchmark over the full evaluation datasets.

\section{Results}
\label{sec:results}
The evaluation separates two questions: how well geometry represents the observed variation within a protein, and how well a fitted model transfers to held-out targets. The first is addressed by within-protein linear fitting, while blind prediction examines the three prediction protocols described above. We first place the results in the context of established flexibility models, then examine matched comparisons, individual structures, and computational scaling.

\subsection{Within-protein fitting and comparison with established flexibility models}
\begin{figure}[htbp]
\centering
\includegraphics[width=0.8\linewidth]{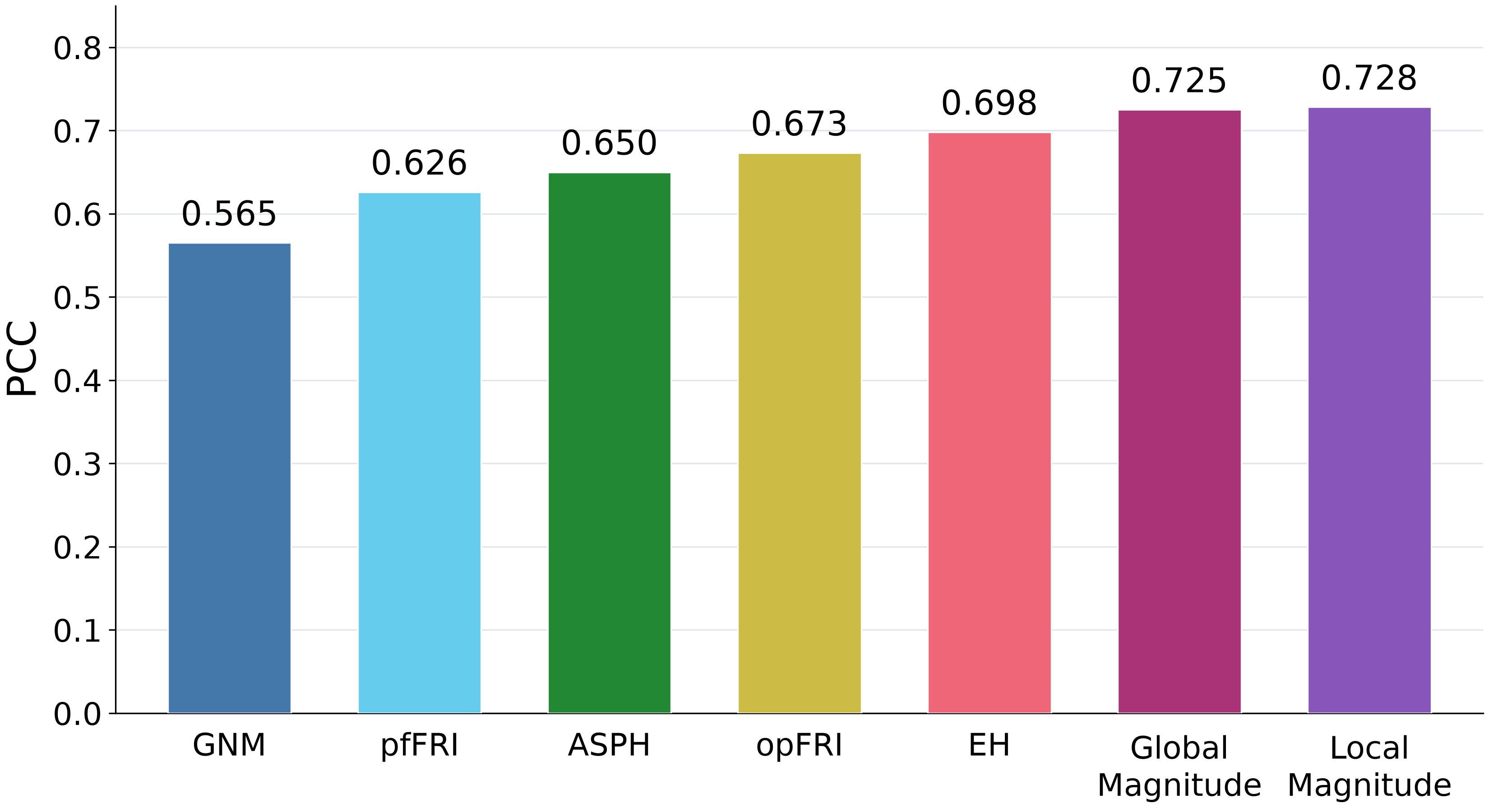}
\caption{Within-protein fitting comparison with established flexibility models. The vertical axis shows the mean per-protein PCC on the Superset. GNM, pfFRI, ASPH, opFRI and EH use mean PCCs from the literature \cite{bi2025} and were not rerun here. Global Magnitude (GMag) and Local Magnitude (LMag) use the global and local 12-feature banks fitted on the same retained coordinates and targets; LMag uses $k=50$.}
\label{fig:literature-comparison}
\end{figure}

Within-protein fitting evaluates how well twelve geometric features describe B-factor variation within each protein, without auxiliary inputs. Figure~\ref{fig:literature-comparison} compares Superset mean PCCs for GNM, parameter-free FRI (pfFRI), atom-specific persistent homology (ASPH), optimized FRI (opFRI), evolutionary homology (EH), and the present global and local magnitude banks. The mean PCCs for the first five methods are reported in \cite{bi2025}. On the aligned Superset, Local Magnitude (LMag) achieves a mean PCC of 0.728, compared with 0.725 for Global Magnitude (GMag); the latter agrees with the reported magnitude PCC of 0.725 at the displayed precision.

Table~\ref{tab:task1} gives the paired global/local results and pooled residue errors. The local bank increases Superset mean PCC by 0.003, while RMSE decreases from 5.242 to 5.181 and MAE from 3.321 to 3.276. These are modest changes in an in-sample fitting task, rather than evidence of a large increase in predictive accuracy.

\begin{table}[htbp]
\centering\small
\caption{Within-protein B-factor fits with 12 features and no auxiliary inputs.
G and L denote the global and local banks. PCC is an equal-weight protein
average; errors pool retained residues.}
\label{tab:task1}
\begin{tabular}{lrrrrrr}
\toprule
&\multicolumn{2}{c}{PCC}&\multicolumn{2}{c}{RMSE ($\Ang^2$)}&\multicolumn{2}{c}{MAE ($\Ang^2$)}\\
\cmidrule(lr){2-3}\cmidrule(lr){4-5}\cmidrule(lr){6-7}
Subset&G&L&G&L&G&L\\
\midrule
Small&0.894&0.888&3.302&3.349&1.850&1.936\\
Medium&0.729&0.729&3.767&3.760&2.570&2.560\\
Large&0.679&0.685&5.435&5.318&2.968&2.895\\
Superset&0.725&0.728&5.242&5.181&3.321&3.276\\
\bottomrule
\end{tabular}
\end{table}

\begin{figure}[htbp]
\centering
\includegraphics[width=\linewidth]{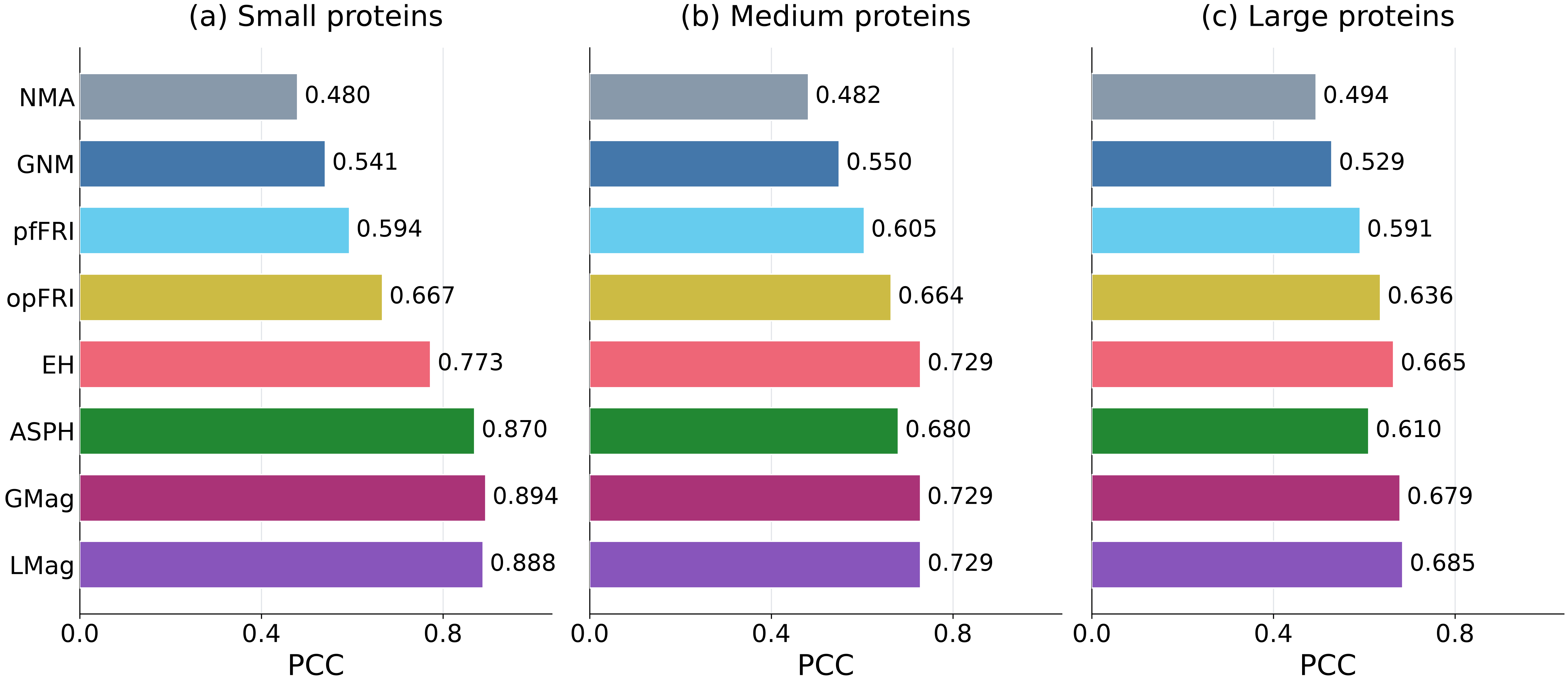}
\caption{Mean per-protein fitting PCC across (a) Small, (b) Medium, and (c) Large protein subsets. NMA, GNM, pfFRI, opFRI, EH, and ASPH; these methods were not rerun here. GMag and LMag are the present paired twelve-feature fits, with $k=50$ for LMag.}
\label{fig:task1-subsets}
\end{figure}

Figure~\ref{fig:task1-subsets} extends the comparison to the supplied Small, Medium, and Large subsets. LMag decreases mean PCC on Small proteins from 0.894 to 0.888, while the small Medium-subset increase leaves both means at 0.729 after rounding. On Large proteins, mean PCC increases from 0.679 to 0.685. The corresponding errors in Table~\ref{tab:task1} change in the same directions. On Small proteins, both magnitude banks exceed the reported ASPH PCC of 0.870. On Medium proteins, all three magnitude results lie close to the reported EH PCC of 0.729. On Large proteins, LMag reaches 0.685, compared with 0.679 for matched GMag, reported PCCs of 0.673 for magnitude and 0.665 for EH.

The size-dependent pattern therefore does not support uniform improvement from localization. Differences in subset membership and coordinate filtering may contribute to why the reported Small/Medium/Large magnitude PCCs do not exactly match the recomputed global values. Comparisons with literature PCCs provide context, while the paired GMag/LMag columns isolate the tested representation change more closely. Even for that paired comparison, the alpha banks differ, so the effects cannot be attributed solely to neighborhood restriction.

\begin{figure}[htbp]
\centering
\includegraphics[width=0.9\linewidth]{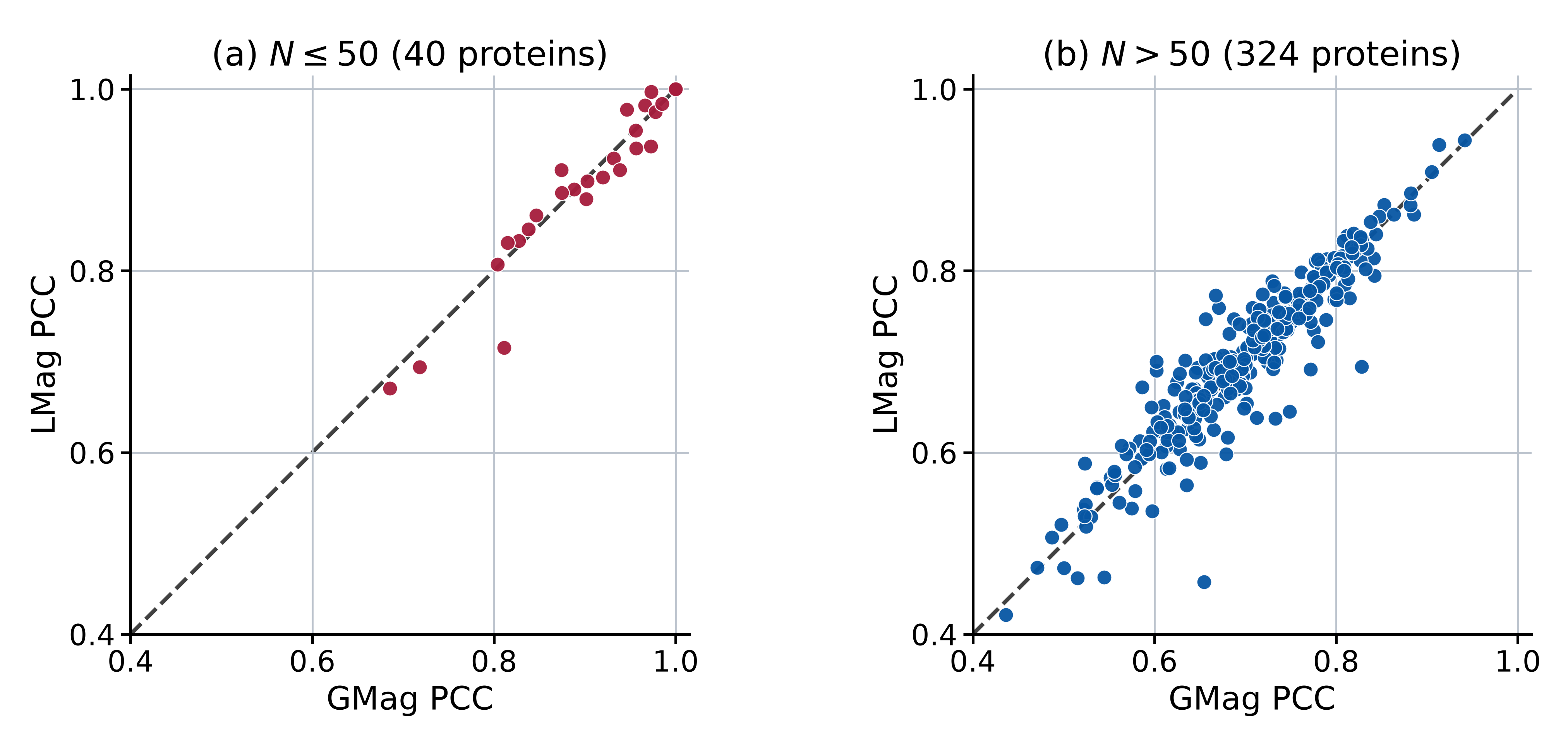}
\caption{Paired Within-protein fitting PCCs stratified by retained C$_\alpha$ count: (a) $N\leq50$ (40 proteins) and (b) $N>50$ (324 proteins). Points above the diagonal favor the local feature bank. Each point represents an in-sample fit; the plot is not a held-out prediction assessment. Coincident points may overlap, particularly near PCC 1.000 in panel (a).}
\label{fig:task1-paired}
\end{figure}

The aggregate gain conceals a distribution of protein-specific effects. With an absolute PCC tolerance of $10^{-8}$ for a numerical tie, the local bank improves 219 proteins, decreases 129, and leaves 16 effectively unchanged. The median paired change is 0.004, with an interquartile range of -0.007 to 0.018. Figure~\ref{fig:task1-paired} shows the paired comparisons separately for $N\leq50$ and $N>50$, using identical axis scales rather than only the cohort mean.

\begin{table}[htbp]
\centering\small
\caption{Within-protein fitting stratified by whether k50 includes the whole structure.
These groups use actual retained C$_\alpha$ counts and differ from the
author-supplied Small, Medium, and Large lists. Ties use $|\Delta|\leq10^{-8}$.}
\label{tab:task1-paired}
\begin{tabular}{lrrrrrr}
\toprule
Group&Proteins&Global PCC&Local PCC&Better&Worse&Tied\\
\midrule
All&364&0.725&0.728&219&129&16\\
$N\leq50$&40&0.933&0.930&11&13&16\\
$N>50$&324&0.699&0.703&208&116&0\\
\bottomrule
\end{tabular}
\end{table}

For $N\leq50$, each local neighborhood contains all retained coordinates. At identical kernel parameters, its center component must equal the global weighting component, up to numerical error. The differences in this group therefore arise from the different alpha banks and numerical fitting, not from truncating the neighborhood. Among the 324 proteins with $N>50$, the mean PCC increases from 0.699 to 0.703. This supports evaluating scale selection and localization separately.

% END GENERATED TASK1 DISTRIBUTION

\subsection{Protein examples: local profiles and size dependence}
\label{sec:protein-examples}
Individual examples connect the weighting construction with fitted B-factor profiles. We use 1ABA, the worked example, together with 3G1S and 1KMM, which also occur in the timing comparison. For a classical network baseline, we additionally compute the Gaussian network model (GNM) on exactly the same retained C$_\alpha$ coordinates \cite{bahar1997}. These examples link structural interpretation, fitting accuracy, and computational cost; they are not an additional validation cohort.

GNM is included because it is a widely used C$_\alpha$ elastic-network model for protein flexibility, providing a classical network-based baseline for comparison with the local Magnitude representation. We use a fixed 7 $\Ang$ contact cutoff and scale its fluctuation amplitudes to B-factor units. The GNM and magnitude fits have different numbers of calibrated coefficients, so their in-sample errors should be interpreted accordingly.

\begin{table}[htbp]
\centering\small
\caption{Within-protein fitting examples with a fixed-7-$\Ang$ GNM baseline recomputed on the same retained coordinates. PCC and RMSE ($\Ang^2$) compare GNM, GMag (G), and LMag (L). GNM amplitudes are scaled to B-factor units; magnitude fits use the within-protein fitting protocol. Examples connect the worked calculation and timing study, and include a local/global PCC decrease.}
\label{tab:protein-examples}
\begin{tabular}{lrrrrrrr}
\toprule
&&\multicolumn{3}{c}{PCC}&\multicolumn{3}{c}{RMSE ($\Ang^2$)}\\
\cmidrule(lr){3-5}\cmidrule(lr){6-8}
Protein&C$_\alpha$ atoms&GNM&G&L&GNM&G&L\\
\midrule
1ABA&87&0.613&0.734&0.752&4.421&2.771&2.691\\
3G1S&418&0.630&0.652&0.647&7.148&6.168&6.201\\
1KMM&1499&0.558&0.670&0.681&13.088&11.522&11.369\\
\bottomrule
\end{tabular}
\end{table}

LMag has a higher PCC than GNM in all three examples, as shown in Table~\ref{tab:protein-examples}, with absolute differences of 0.139, 0.017, and 0.123 for 1ABA, 3G1S, and 1KMM, respectively. For 3G1S, localization slightly reduces PCC relative to GMag and increases RMSE, whereas 1KMM improves in both measures. Thus increasing structure size does not guarantee a local/global gain for every protein. The 1KMM feature calculation also benefits from bounded neighborhoods (Section~\ref{sec:timing-results}), but fitting accuracy and computation time remain distinct properties. The GNM comparisons here concern these three structures; they do not replace the literature cohort summaries with a new full-cohort GNM evaluation.

\begin{figure}[htbp]
\centering
\includegraphics[width=\linewidth]{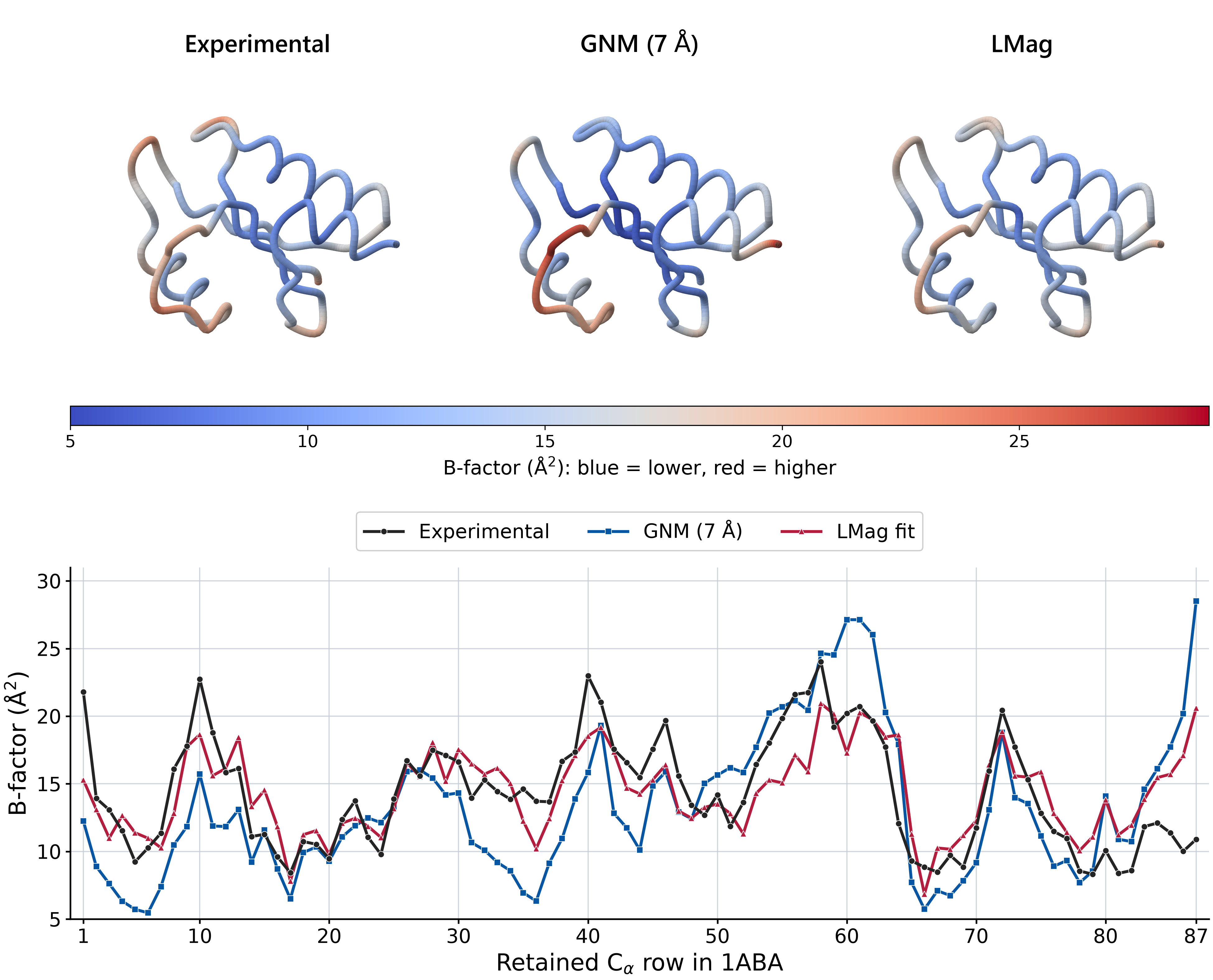}
\caption{Structural colouring and B-factor profiles for 1ABA. Upper panels: the same 87-C$_\alpha$ backbone coloured by experimental B-factors, fixed-7-$\Ang$ GNM (PCC 0.613), and LMag (PCC 0.752), respectively. A common blue-to-red scale indicates increasing B-factor. Lower panel: experimental values (black circles), GNM (blue squares), and LMag fit (red triangles). The horizontal axis is retained row index, not an assumed uninterrupted residue numbering.}
\label{fig:profile}
\end{figure}

For the illustrative protein 1ABA, LMag has PCC 0.752, compared with 0.613 for the fixed-7-$\Ang$ GNM; the global magnitude fit has PCC 0.734. Figure~\ref{fig:profile} compares the experimental B-factors with GNM and LMag, using exactly the retained row order. The upper panels share the same geometry, view, and colour range, so their differences reflect experimental or modeled B-factors rather than structural deformation. GNM provides the classical network baseline; LMag uses the twelve-feature within-protein fit. This example is not an independent test or a substitute for the full-cohort summary.

% END GENERATED PROFILE

The 1ABA profiles capture broad variations in the experimental B-factors, although residue-level discrepancies remain. A higher overall PCC therefore does not imply an accurate fit at every residue.

\subsection{Blind prediction results and analysis}
\label{sec:task2}
Blind prediction evaluates held-out B-factor prediction using twelve magnitude features and the same 37 auxiliary descriptors for both representations. Protein ten-fold and leave-one-protein-out (LOPO) hold out entire proteins, whereas atom ten-fold holds out residues and can include other residues from the same protein in training; it therefore addresses a different prediction setting.

\begin{figure}[htbp]
\centering
\includegraphics[width=0.8\linewidth]{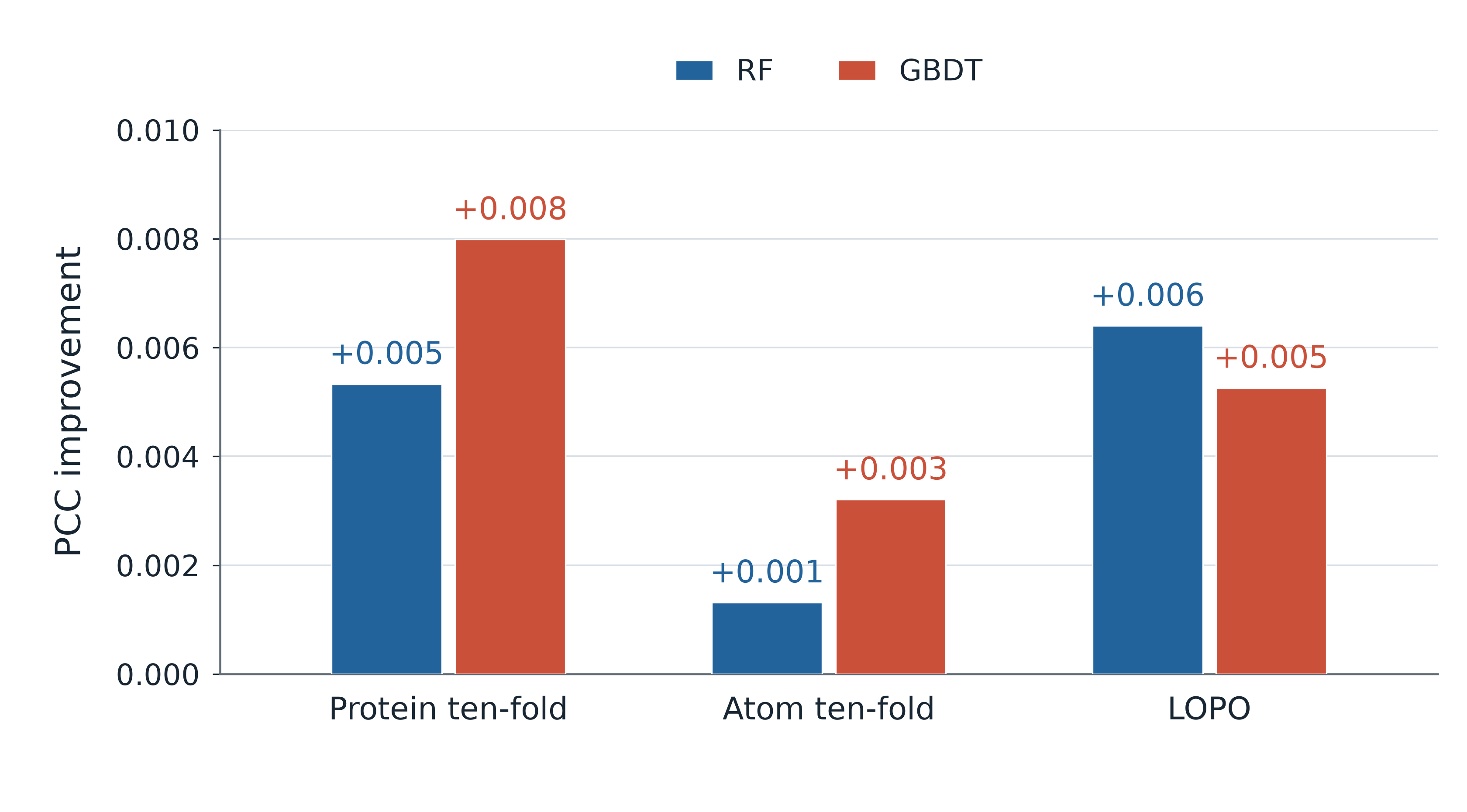}
\caption{Absolute improvements in the primary blind-prediction PCC for a fixed model
seed. Bars show LMag minus GMag PCC for RF and GBDT under protein ten-fold,
atom ten-fold and leave-one-protein-out (LOPO) evaluation.}
\label{fig:pcc}
\end{figure}

Table~\ref{tab:task2pcc} compares PCCs reported in the literature with those from the controlled global/local comparison. Across the six model--protocol combinations, the local representation increases the primary PCC in every case. Absolute differences range from 0.001 for RF atom ten-fold to 0.008 for GBDT protein ten-fold, as shown in Figure~\ref{fig:pcc}. The corresponding RMSE and MAE also decrease in all six comparisons in Table~\ref{tab:errors}.

\begin{table}[htbp]
\centering\small
\caption{Blind-prediction PCC: literature values \cite{bi2025} versus same-data global and local
49-dimensional inputs. The cited study uses repeated evaluations; the last three
columns use one fixed seed. $\Delta$ is local minus global. Ten-fold and
LOPO aggregation conventions are defined in Section~\ref{sec:evaluation}.}
\label{tab:task2pcc}
\begin{tabular}{llrrrr}
\toprule
Protocol&Model&Literature PCC&Global49&Local49&$\Delta$\\
\midrule
Protein ten-fold&RF&0.386&0.383&0.388&0.005\\
Protein ten-fold&GBDT&0.424&0.433&0.441&0.008\\
Atom ten-fold&RF&0.841&0.842&0.843&0.001\\
Atom ten-fold&GBDT&0.836&0.836&0.839&0.003\\
LOPO&RF&0.531&0.531&0.538&0.006\\
LOPO&GBDT&0.570&0.570&0.575&0.005\\
\bottomrule
\end{tabular}
\end{table}

GBDT performs better than RF under the protein-held-out protocols, whereas RF is stronger in atom ten-fold PCC. The latter protocol yields higher correlations overall, but this cannot be interpreted as superior prediction for unseen proteins: residues from the same protein can inform the training set.
\begin{table}[htbp]
\centering\small
\caption{Pooled blind-prediction absolute errors, in $\Ang^2$. The last column is RMSE
of the training-fold mean predictor, identical for both feature banks.}
\label{tab:errors}
\begin{tabular}{llrrrrr}
\toprule
&&\multicolumn{2}{c}{RMSE}&\multicolumn{2}{c}{MAE}&Mean baseline\\
\cmidrule(lr){3-4}\cmidrule(lr){5-6}
Protocol&Model&Global&Local&Global&Local&RMSE\\
\midrule
Protein ten-fold&RF&13.200&13.178&7.824&7.799&11.949\\
Protein ten-fold&GBDT&12.425&12.404&7.499&7.470&11.949\\
Atom ten-fold&RF&6.301&6.277&4.318&4.298&11.663\\
Atom ten-fold&GBDT&6.666&6.620&4.627&4.593&11.663\\
LOPO&RF&13.165&13.148&7.745&7.721&11.938\\
LOPO&GBDT&12.436&12.432&7.512&7.506&11.938\\
\bottomrule
\end{tabular}
\end{table}

The aggregate improvements conceal variation across proteins. In LOPO, PCC increases for 250 of 345 proteins with RF and for 238 of 345 with GBDT. Nevertheless, GBDT's Small-subset PCC decreases from 0.436 to 0.432, see Table~\ref{tab:lopo}, so the gain is not uniform across subgroups.

\begin{table}[htbp]
\centering\small
\caption{LOPO subgroup PCCs on the retained blind-prediction cohort. G and L use the same
auxiliary features. Literature PCCs \cite{bi2025} are shown separately.}
\label{tab:lopo}
\begin{tabular}{lrrrrrrr}
\toprule
&&\multicolumn{3}{c}{RF}&\multicolumn{3}{c}{GBDT}\\
\cmidrule(lr){3-5}\cmidrule(lr){6-8}
Subset&Proteins&Literature PCC&G&L&Literature PCC&G&L\\
\midrule
Small&30&0.390&0.388&0.398&0.437&0.436&0.432\\
Medium&37&0.530&0.528&0.529&0.550&0.551&0.554\\
Large&34&0.518&0.516&0.520&0.565&0.565&0.569\\
Superset&345&0.531&0.531&0.538&0.570&0.570&0.575\\
\bottomrule
\end{tabular}
\end{table}

The gains from localization are modest and do not extend to every subgroup. Moreover, under protein-held-out evaluation, RMSE remains above the training-mean baseline, despite improvements over the global representation. Better correlation should therefore not be interpreted as accurate cross-protein B-factor calibration. These blind-prediction results also reflect the contribution of the 37 auxiliary descriptors shared by both models.

\subsection{Computational scaling on individual protein examples}
\label{sec:timing-results}
Both feature extractors were measured on an Intel Core Ultra 9 275HX under Windows with a single process and one BLAS thread. Table \ref{tab:timing} reports three-repeat medians from the five-protein reproducibility run; the accompanying evidence retains each measurement. These times are specific to the stated machine and implementation.

\begin{table}[htbp]
\centering
\caption{Five real-protein timing examples. Seconds are medians of three
repeats. Both methods compute all 12 features with identical alpha values
and one BLAS thread. Global/local greater than one favors local.}
\label{tab:timing}
\begin{tabular}{lrrrr}
\toprule
Protein&C$_\alpha$ atoms&Global (s)&Local (s)&Global/local\\
\midrule
1ABA&87&0.001&0.046&0.020\\
3G1S&418&0.061&0.228&0.270\\
1KMM&1499&2.252&0.813&2.770\\
1H6V&2927&12.485&1.670&7.480\\
1QKI&3912&28.925&2.194&13.180\\
\bottomrule
\end{tabular}
\end{table}

\begin{figure}[htbp]
\centering
\includegraphics[width=0.8\linewidth]{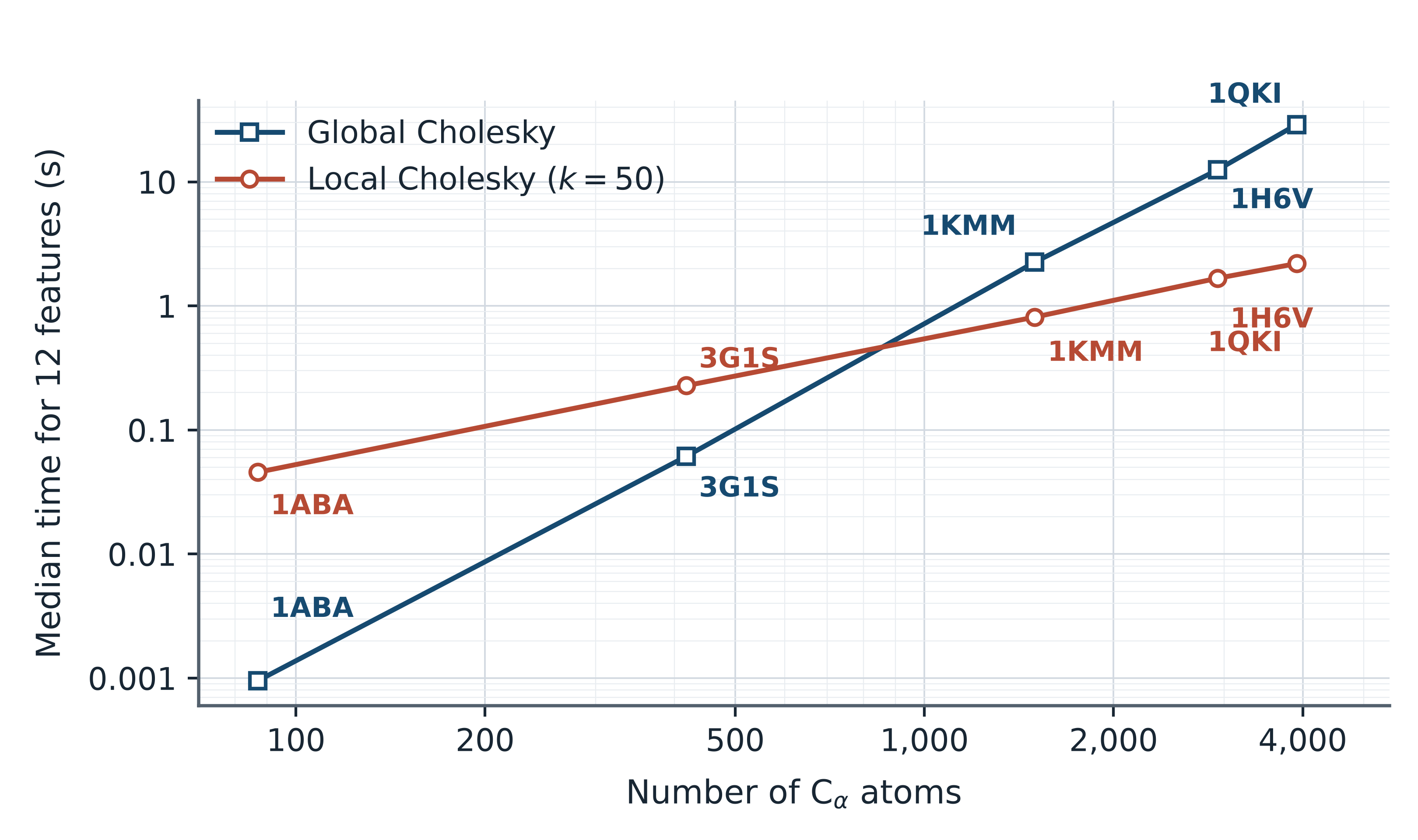}
\caption{Measured size dependence for the five examples in
Table~\ref{tab:timing}. Connecting lines guide the eye; they are not a fitted
complexity model. The large-structure advantage does not imply a speedup
on collections dominated by small proteins.}
\label{fig:timing}
\end{figure}

The global solver is faster on the smallest examples, because it obtains all weights with only 12 factorizations. At larger $N$, the bounded local factorizations become advantageous, as predicted by \eqref{eq:complexity}. The 3912-C$_\alpha$ structure 1QKI is included explicitly rather than extrapolated from smaller point clouds. No universal crossover size is inferred: implementation details, CPU state, threading, conditioning, and geometry influence the observed ratios. Also, these are features from a structure identified by a PDB accession and need not represent a single chain.

The measured examples therefore indicate that local mgnitude feature extraction offers no speed advantage for small protein structures, but becomes faster than global extraction for large structures.

\FloatBarrier
\section{Discussion and conclusion}
\label{sec:discussion}
\subsection{Effective local features for B-factor modeling}
This study demonstrates that a fixed local magnitude representation retains useful structural information for protein B-factor modeling. Each C$_\alpha$ atom is described by twelve central weighting components computed from at most 50 neighboring atoms. Because the kernel solve incorporates interactions among the neighbors, the descriptor captures their spatial arrangement in addition to their distances from the center. This provides a compact description of the local structural environment.

On the evaluated cohorts, the local representation increases the mean within-protein fitting PCC from 0.725 to 0.728 and improves the primary PCC in all six blind-prediction model--protocol comparisons. These results establish that the selected local feature bank remains effective despite restricting the structural context. The gains are modest and are not uniform across protein subsets, with lower mean PCC on the Small subsets. In blind prediction, both representations are combined with the same 37 auxiliary descriptors; the comparison therefore measures the effect of replacing the geometric feature block within a common prediction model.

The principal value of local magnitude lies in retaining informative structural features while achieving a practical reduction in feature-extraction cost for large protein structures. This conclusion is supported by the predictive comparisons and the separate timing experiments. It does not require local weights to reproduce global weights: the local representation is evaluated through its ability to describe B-factor variation.

\subsection{Efficiency and outlook}
For a fixed twelve-feature bank, dense global factorization has leading cost $O(N^3)$, whereas local factorization costs $O(Nk^3)$. With $k=50$, the local factorization cost grows linearly with the number of centers. The five measured protein examples confirm the expected large-structure advantage, while showing that repeated small solves can make local extraction slower on small structures. Local magnitude therefore offers a practical route to reducing the cost of geometric feature extraction as protein size increases.

The predictive interpretation remains specific to the present configuration. Global and local features use different scale banks, parameters were informed by exploration on overlapping proteins, and the experiments use one model seed. A matched-scale comparison with training-only parameter selection, repeated seeds, and independent protein-level validation is needed to determine how much of the observed gain arises from localization. Proteins with $N\leq50$ provide a useful control: their local and global center weights coincide for identical kernels, up to numerical error.

Future work should identify the smallest neighborhood that preserves predictive performance under a specified computational budget. Comparing neighborhood sizes at matched scales, and relating the resulting errors to protein size and local density, would help determine where additional structural context is useful. Adaptive neighborhoods are a further possibility, but any selection rule should be established using training data and assessed on held-out proteins. The objective is an efficient representation with reliable predictive information, rather than a uniformly larger neighborhood.

In conclusion, local magnitude provides a compact, geometry-based representation that retains effective B-factor features and reduces feature-extraction time on the large protein examples examined here. Its main contribution is the combination of useful predictive information and improved computational scalability, with the strongest practical motivation arising in applications involving large structures.

\section*{Data and code availability}
The data and code supporting this study is available at \url{https://github.com/Wanyingbear/local_magnitude_algorithm}.

\appendix
\section{Stability for a fixed neighborhood}
\label{app:stability}
Fix an ordered neighborhood $S_i=(s_1,\ldots,s_m)$ with center $s_1=i$ and kernel parameters $\alpha>0$, $p\in\{1,2\}$. Let $A$ and $\widetilde A$ be the kernels in \eqref{eq:kernel} evaluated on corresponding original and perturbed coordinates, with the same indices and ordering. Assume that the points in each configuration are pairwise distinct. Both kernels are then symmetric positive definite, independently of any small-perturbation assumption. Set $E=\widetilde A-A$; this difference need not be positive definite. Throughout, $\norm{\cdot}_2$ denotes the Euclidean norm for vectors and the induced spectral norm for matrices; $\one_m$ is the all-ones vector, and $e_1$ is the first coordinate vector. Define
\begin{equation*}
 w=A^{-1}\one_m,\quad \widetilde w=\widetilde A^{-1}\one_m,\quad f_i=e_1^Tw,\quad \Mag_i=\one_m^Tw,
\end{equation*}
with $\widetilde f_i$ and $\widetilde{\Mag}_i$ defined analogously from $\widetilde w$.

\begin{proposition}
Let $m>1$ and suppose each coordinate moves by at most $\varepsilon$ in Euclidean norm. Set $L_{\alpha,1}=\alpha$ and $L_{\alpha,2}=\sqrt{2\alpha/e}$. Then
\begin{equation}
 \norm{E}_2\leq2(m-1)L_{\alpha,p}\varepsilon.
 \label{eq:stability-coordinates}
\end{equation}
In particular, if
\begin{equation}
 \varepsilon<\frac{\lambda_{\min}(A)}{2(m-1)L_{\alpha,p}},
 \label{eq:stability-threshold}
\end{equation}
then $\eta=\norm{A^{-1}}_2\norm{E}_2<1$. With $B_E=\sqrt m\,\norm{A^{-1}}_2^2\norm{E}_2/(1-\eta)$,
\begin{equation}
 \norm{\widetilde w-w}_2\leq B_E,\qquad
 |\widetilde f_i-f_i|\leq B_E,\qquad
 |\widetilde{\Mag}_i-\Mag_i|\leq\sqrt m\,B_E.
 \label{eq:stability-bounds}
\end{equation}
\end{proposition}

\begin{proof}
Write $d_{ab}=\norm{x_{s_a}-x_{s_b}}_2$ and $\widetilde d_{ab}=\norm{\widetilde x_{s_a}-\widetilde x_{s_b}}_2$. The reverse triangle inequality and the coordinate perturbation assumption give
\begin{equation*}
 |\widetilde d_{ab}-d_{ab}|
 \leq \norm{\widetilde x_{s_a}-x_{s_a}}_2
      +\norm{\widetilde x_{s_b}-x_{s_b}}_2
 \leq 2\varepsilon.
\end{equation*}
For $\phi_p(r)=e^{-\alpha r^p}$, we have $|\phi_1'(r)|=\alpha e^{-\alpha r}\leq\alpha$ and $|\phi_2'(r)|=2\alpha r e^{-\alpha r^2}\leq\sqrt{2\alpha/e}$; the latter maximum occurs at $r=(2\alpha)^{-1/2}$. The mean-value inequality therefore yields
\begin{equation*}
 |E_{ab}|=|\phi_p(\widetilde d_{ab})-\phi_p(d_{ab})|
 \leq L_{\alpha,p}|\widetilde d_{ab}-d_{ab}|
 \leq 2L_{\alpha,p}\varepsilon.
\end{equation*}
Both kernels have unit diagonal, so $E_{aa}=0$. Since $E$ is symmetric, its spectral norm is bounded by the largest absolute row sum
\begin{equation*}
 \norm{E}_2\leq\max_{1\leq a\leq m}\sum_{b=1}^m|E_{ab}|
 \leq2(m-1)L_{\alpha,p}\varepsilon.
\end{equation*}
Positive definiteness of $A$ gives $\norm{A^{-1}}_2=1/\lambda_{\min}(A)$, hence \eqref{eq:stability-threshold} implies
\begin{equation*}
 \eta=\norm{A^{-1}}_2\norm{E}_2
 \leq\frac{2(m-1)L_{\alpha,p}\varepsilon}{\lambda_{\min}(A)}<1.
\end{equation*}

To bound the inverse, factor $\widetilde A=A(I_m+A^{-1}E)$, where $I_m$ is the identity matrix. Since $\norm{A^{-1}E}_2\leq\eta<1$, the Neumann series for $(I_m+A^{-1}E)^{-1}$ converges and gives
\begin{equation*}
 \norm{\widetilde A^{-1}}_2
 \leq\left(\sum_{j=0}^{\infty}\eta^j\right)\norm{A^{-1}}_2
 =\frac{\norm{A^{-1}}_2}{1-\eta}.
\end{equation*}
The equations $Aw=\one_m$ and $\widetilde A\widetilde w=\one_m$ imply $\widetilde A(\widetilde w-w)=-Ew$. Using $\norm{w}_2\leq\norm{A^{-1}}_2\norm{\one_m}_2=\sqrt m\,\norm{A^{-1}}_2$, we obtain
\begin{equation*}
 \norm{\widetilde w-w}_2
 \leq\norm{\widetilde A^{-1}}_2\norm{E}_2\norm{w}_2
 \leq\frac{\sqrt m\,\norm{A^{-1}}_2^2\norm{E}_2}{1-\eta}=B_E.
\end{equation*}
Finally, Cauchy--Schwarz gives $|\widetilde f_i-f_i|=|e_1^T(\widetilde w-w)|\leq\norm{\widetilde w-w}_2$ and $|\widetilde{\Mag}_i-\Mag_i|=|\one_m^T(\widetilde w-w)|\leq\sqrt m\,\norm{\widetilde w-w}_2$, proving \eqref{eq:stability-bounds}.
\end{proof}

Thus, for each fixed kernel $A$, sufficiently small coordinate perturbations ensure $\eta<1$ and local Lipschitz continuity of the weights and neighborhood magnitude. Positive definiteness guarantees that both weightings exist. The small-perturbation condition supplies a quantitative error bound. These bounds establish local stability for fixed neighborhoods, with sensitivity governed by the smallest eigenvalue of the kernel matrix.

\section{Model and feature specifications}
\label{app:models}
\begin{table}[htbp]
\centering\small
\caption{Fixed blind-prediction model settings. Both methods use the same values.}
\begin{tabular}{lll}
\toprule
Parameter&RF&GBDT\\
\midrule
Number of trees&1000&1000\\
Maximum depth&8&7\\
Minimum split size&4&5\\
Minimum leaf size&2&1 (library default)\\
Features considered per split&All&Square root of input count\\
Learning rate&Not applicable&0.002\\
Subsample fraction&Bootstrap default&0.8\\
Random seed&0&0\\
Feature/target standardization&None&None\\
\bottomrule
\end{tabular}
\end{table}

\begin{table}[htbp]
\centering\small
\caption{The 49-dimensional blind-prediction input. Column ranges are one-based.}
\begin{tabularx}{0.8\linewidth}{lrX}
\toprule
Columns&Count&Meaning\\
\midrule
1--12&12&Local or global weighting features, ordered by p and then alpha\\
13--32&20&Amino-acid one-hot indicators\\
33--39&7&Secondary-structure one-hot indicators\\
40--49&10&$\phi$, $\psi$, area, resolution, R-value, heavy-atom count,
occupancy, three packing-density descriptors\\
\bottomrule
\end{tabularx}
\end{table}

The global and local timing implementations use double-precision Cholesky without diagonal jitter. Tree estimators may convert inputs internally to single precision; the reference experiment checked that reused global inputs agreed exactly after that conversion. Auxiliary descriptors were supplied by the original dataset, and this study does not provide a complete auxiliary-feature generation pipeline for arbitrary new structures.

\section*{Acknowledgments}

This work was supported in part by the Natural Science Foundation of China [Grant Number 12401080], the Scientific Research Foundation of Chongqing University of Technology, the Start-up Research Fund of the University of North Carolina at Charlotte, and the University of North Carolina at Charlotte KCOS Faculty Research Grant.

\end{document}